\documentstyle[12pt,epsf]{article}

\begin{document}        
\pagestyle{empty}
\renewcommand{\thefootnote}{\fnsymbol{footnote}}

\begin{flushright}
{\small
SLAC--PUB--8773\\
Feb 2001\\}
\end{flushright}
 
\vspace{.8cm}


\begin{center}
{\bf\large   
Study of Top-quark Production and Decay Vertices with LCD Fast
Simulation
\footnote{Work supported 
by Department of Energy contract  DE--AC03--76SF00515 (SLAC).}}

\vspace{1cm}
{\bf
Masako Iwasaki
} \\
\vskip 0.2cm
{\it
Department of Physics, University of Oregon, Eugene,
OR 97403-1274, USA
} 

\medskip
\end{center}
 
\vfill

\begin{center}
{\bf\large   
Abstract }
\end{center}
We report 
a study of top-quark reconstruction in $e^+e^- \rightarrow t\bar{t}$
events at a 500 GeV linear collider using the LCD Fast simulator.
The initial study of top-quark anomalous couplings is also reported.
The final states of 4 jets and lepton as well as 6 jets are used.
Using the 4 jets and lepton final state, we estimate
the preliminary sensitivities for form factors at 
the $\gamma/Z^0 t\bar{t}$ vertex.
In the 6 jets reconstruction, we show abilities of 
the top-quark charge identification and the $c$-quark tagging 
in W decays.
\vfill

\begin{center} 
{\it Presented at the 5th International Linear Collider Workshop 
(LCWS 2000), 24-28 Oct 2000, Fermilab, Batavia, Illinois, USA} 
\end{center}

\newpage
\pagestyle{plain}

\section{Introduction}
The top quark plays a special role in particle physics, 
due to its uniquely large mass\cite{CDF D0}. 
It decays immediately, before forming a hadron.
Therefore its spin information is expected to be directly 
transfered to its daughters.
This feature provides us opportunities of probing the top couplings 
using the information of its daughters.
The high energy future $e^+e^-$ linear collider would be an ideal
tool for such studies, because of 
its clean event environment and 
the possibility of initial-state polarization.

In this report, we will give a study of top-quark reconstruction 
in $e^+e^- \rightarrow t\bar{t}$, and the initial results of 
the top-quark anomalous coupling analysis.
There are three kinds of final states in the $t\bar{t}$ production:
(i)two $b$ jets and four jets from W's (45\%), 
(ii)two $b$ jets, two jets and one charged lepton(44\%), or 
(iii)two $b$ jets and two charged leptons(11\%).
We study cases (i) and (ii),
where (i) six jets and (ii) four jets and one charged lepton
are in the final states.
In the anomalous coupling analysis, 
we use the same notations of general top-quark couplings
and angular definitions of polar angle, 
top-decay angle and W-decay angle, as Ref.\cite{Schmidt}.

\section{Event Analysis}
We generate 60,000 $e^+e^- \rightarrow t\bar{t}$ events
corresponding to an integrated luminosity of 80 $fb^{-1}$ 
with the PANDORA program\cite{pandora}, with parton showers and 
hadronization done by PYTHIA 6.1\cite{pythia}.
A top-quark mass of 175 GeV, a beam energy of 250 GeV and 
80\% left-handed electron polarization are assumed.
Beamstrahlung and initial state radiation are included.
No other final states are generated.

To simulate the detector, we use the 
Linear Collider Detector (LCD) Fast simulation, 
assuming the Large design\cite{LCD param}.
Charged tracks with E$_{track} > 200$ MeV and 
$|\cos\theta|<0.90$, and clusters 
with E$_{cluster} > 300$ MeV and $|\cos\theta|<0.98$
are used.

\subsection {4 Jets and Lepton Analysis}
In this analysis, we use only $t\bar{t}\rightarrow$ 4 jets $+$
lepton (muon or electron) events (17343 events).
We apply the event-selection cuts of 
i) the number of charged tracks $\geq$ 20, 
ii) there is a lepton track (muon or electron) with momentum $>$ 20 GeV, 
and iii) the visible energy $>$ 300 GeV.
Visible energy is calculated
with charged tracks and neutral clusters.
No other criteria for lepton identification
are required.

Jets are reconstructed with ``energy flow'' objects, 
consisting of charged tracks and neutral clusters.
Neutral clusters are selected by
the absence of a track and cluster association.
Here all charged tracks are extrapolated to the cluster cylindrical
radius.
Then the clusters which have any track with track-cluster 
distance $<$ 8cm are regarded as charged clusters. 
With this cut, we reject 
93\% charged while keeping 92\% neutral clusters.
Using the energy flow objects, except for the charged lepton,
we reconstruct jets using an invariant-mass (JADE) algorithm.
First we apply $Y_{\rm cut} = 0.008$ and select events 
which have 4 or more jets. Then the $Y_{\rm cut}$ value is increased,
if necessary, until the event has exactly 4 jets.
The efficiency of this selection  is 57\% for $t\bar{t} \rightarrow$ 
4 jets events.

To tag the $b$ jets, we use the method developed by 
SLD\cite{ZVTOP}. The secondary vertex is topologically 
reconstructed with charged tracks, 
and its $P_T$-corrected mass ($M_{P_T}$) is calculated.
The jets with $M_{P_T}$ $>$ 1.8 GeV are tagged as $b$ jets.
The efficiency and purity of 61\% and 95\% are obtained, respectively.
To identify the light-flavor ($uds$) jets, we use the $N_{\rm sig}$
method,
where $N_{\rm sig}$ is the number of tracks which have 3D impact 
parameter with significance $> 3 \sigma$ (excluding $V^0$ decay tracks). 
We select $uds$ jets by requiring $N_{sig} = 0$ with 87\% efficiency 
and 77\% purity.

To reconstruct $W$'s, we search all two-jets combinations.
Here we apply the flavor tagging that 1) none of two jets is 
tagged as a $b$ jet, and 2) at least one jet is tagged as a $uds$ jet.
Jet pairs with invariant mass within 15 GeV of 
the nominal W mass are kept. 
Top-quark candidates are then formed from these W's and $b$ jets. 
To reduce random combinatoric background,
we require $x_E \equiv E_{3jets}/E_{beam}$ satisfies
$0.9 < x_E < 1.1$.
The combinations with invariant mass in the range 
160 GeV to 190 GeV are regarded as top-quark candidates.
We select 2295 (5731 for without flavor tagging) top-quark 
candidates, with purity of 90\% (84\%).
In this sample, 89\% (56\%) of the candidates have the
correct $b$ and W assignment.
The mass resolution for the reconstructed top quark is 7.8 GeV (9.2 GeV).

\begin{figure} 
\begin{center}
\epsfysize 2.1in
\epsfbox{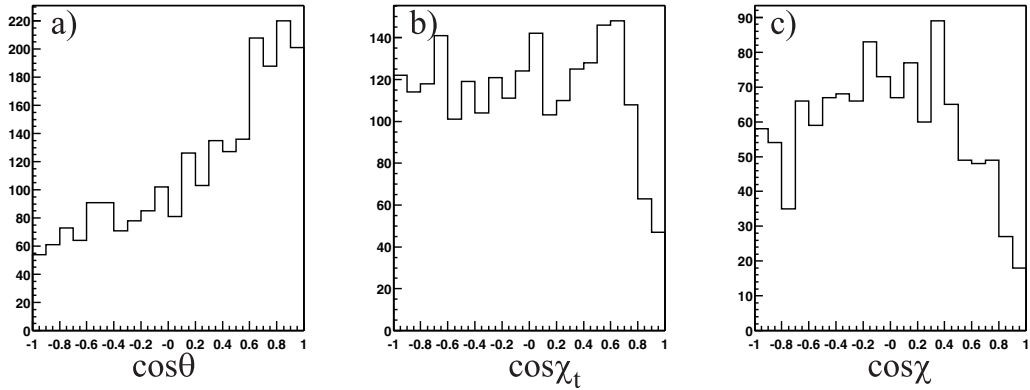}
\end{center}
\caption{The angular distributions of
a) $\cos\theta$, b) $\cos\chi_t$ and c) $\cos\chi$
for the reconstructed $t\bar{t}$ candidates. The definitions of the
$\chi_t$ and $\chi$ are in the text.}
\label{fig:angle}
\end{figure}

Then we reconstruct the polar angle ($\theta$) of top quark, 
top-decay angle ($\chi_t$) and W-decay angle ($\chi$). 
The charge of the top quark is determined by the charge
of the lepton.
In the 4 jets and lepton analysis, we only reconstruct one
$t$($\bar{t}$) quark, which has 3 jets in the final state.
Since the final state of the other $\bar{t}(t)$ quark includes 
a neutrino which can not be detected, we assume that the
$\bar{t}(t)$ quark has the same energy and the opposite momentum 
as the reconstructed $t(\bar{t})$ quark.
We also assume that
1) $b$ jet and W are back-to-back in the top rest frame, and
2) lepton and neutrino are back-to-back in the W rest frame,
to reconstruct the top-decay and W-decay angles.
Fig.~\ref{fig:angle} shows the reconstructed angular distributions.
Reconstruction efficiencies in $\cos\chi_t{>}0.6$ 
and $cos\chi{<}0$ are significantly dropped because of the acceptance cut.
The angular resolutions are 35 mrad, 69 mrad and 
115 mrad for $\theta$, $\chi_t$ and  $\chi$, respectively.

Using the angular distributions, 
we estimate the preliminary sensitivities 
of 0.023 and 0.034 (normalized)
for $F_{1A}^\gamma$ and $F_{1A}^Z$, respectively, 
at the $\gamma/Z^0 t\bar{t}$ vertex 
for an integrated luminosity of
80 $fb^{-1}$ with 80\% left-handed electron beams.
%

\subsection {6 Jets Analysis}
In this study, we use 27730 $t\bar{t}\rightarrow 6$ jets
events in the generated $t\bar{t}$ sample described above.
For the event selection, we require 
i) the number of charged tracks $\geq$ 30
and ii) visible energy $>$ 350 GeV.
The energy flow objects are used to form jets.
As before, we first apply $Y_{cut} = 0.003$ to select events 
with 6 or more jets, then increase $Y_{cut}$ until the event 
has exactly 6 jets.
We require that an event has exactly 2 $b$ jets, 
where $b$ jets are identified by a cut $M_{P_T}$ $>$ 1.8GeV.
After identifying the two $b$ jets, we exclusively reconstructed
two W's from the remaining 4 jets.
Top-quark candidates are then formed from W's and $b$ jets.
We require at least one W candidate has an invariant mass 
within 15 GeV of the nominal W mass, and at least one top 
candidate satisfies 160 GeV $<$ $M_{top}$ $<$ 190 GeV.

To identify the charge of the top quarks, we measure 
Vertex Charge of $b$ jets.
Here the Vertex Charge is a charge sum of the 
secondary vertex reconstructed by the topological vertexing.
We require at least one of two $b$ jets has the non-zero Vertex-Charge
value. We discard the events which have the same-sign $b$ jets.
The efficiency of this cut is 57\%, 
and the charge of the top quark is determined with 83\% purity.
The provability of the correct assignment of $b$ and W in the top-quark
candidates is 78\%.
We reconstruct the top-quark polar angle $\theta$
and the top-decay angle $\chi_t$ with resolutions of $\Delta\theta$ = 48 mrad 
and $\Delta \chi_t$ = 88 mrad.

To reconstruct the W-decay angle, 
we tag the $c$ quark in the W decay 
using the $P_T$-corrected mass ($M_{P_T}$) 
and momentum of the secondary vertex ($P_{VTX}$).
We identify the $c$ jets with the criteria 
i) $20 \times M_{P_T} - P_{VTX} < 10$ 
ii) 0.6 GeV $< M_{P_T} <$ 1.8 GeV, and
iii) $P_{VTX} >$ 10 GeV.
The purity and efficiency for $c$ quarks are 98\% and 33\%, respectively.
Here we get the high purity because there is little $b$ background 
in the reconstructed W's.
We select 689 $c$-jet candidates, and reconstruct the W-decay angles
with the angular resolution of 90 mrad.

Since we use the heavy-flavor tagging and the charge identification
with the Vertex Charge in this analysis, 
the Vertex Detector performance is 
expected to be important.
Changing the inner radius of the Vertex
Detector from 1cm to 2cm, we lose 1\% of 
the top-quark reconstruction efficiency
and 13\% of the $c$-quark tagging efficiency in W decays. 
Therefore the Vertex Detector 
performance is important especially for the $c$-quark tagging.

\section{Conclusion}
We have studied the top-quark reconstruction in $e^+e^- \rightarrow t\bar{t}$
events at a 500 GeV linear collider using the LCD Fast simulator. 

In the 4 jets and lepton analysis, we estimate the preliminary
sensitivities of $F_{1A}^{\gamma}$ and $F_{1A}^{Z}$ 
at the $\gamma/Z^0 t\bar{t}$ vertex 
for an integrated luminosity of
80 $fb^{-1}$ with 80\% left-handed electron beams.
To estimate the other coupling sensitivities, 
the detailed acceptance correction
and background studies are necessary.

In the 6 jets analysis, we determine the charge of the top quark using
the Vertex Charge of $b$ jet with 83\% purity. 
We also apply the $c$-quark tagging in W decays.
The Vertex Detector performance is important especially for the 
$c$-quark tagging.


\begin{thebibliography}{99}

\bibitem{CDF D0}
CDF collaboration, T. Affolder {\em et al.},
Phys. Rev. {\bf D63}~032003~(2001);\\
D0 collaboration, B.~Abbott {\em et al.}, 
Phys. Rev. {\bf D58}~052001~(1998). 
\bibitem{Schmidt}
C. Schmidt, Phys. Rev. {\bf D54}~3250~(1996).
\bibitem{pandora} 
M. Peskin, hep-ph/9910519 (1999).
\bibitem{pythia} 
T. Sj\"ostrand, Comp. Phys Comm. {\bf 82}~74~(1994).
\bibitem{LCD param} 
M. Iwasaki and T. Abe, 
``LCD ROOT Simulation and Analysis Tools'', These proceedings. 
\bibitem{ZVTOP} 
D. Jackson, Nucl. Inst. Meth. {\bf A388}~247~(1997).
\end{thebibliography}
\end{document}